\documentclass[aps,prl,reprint,twocolumn,amsmath,amssymb,showpacs,superscriptaddress]{revtex4-1}

\usepackage{color}
\usepackage{textcomp} 
\usepackage{mathrsfs,amsmath}
\usepackage{graphicx}
\usepackage{lipsum}
\usepackage{dcolumn}
\usepackage[mathlines]{lineno}
\usepackage{hyperref}
\usepackage{xcolor}
\hypersetup{
    colorlinks,
    linkcolor={red!50!black},
    citecolor={blue!50!black},
    urlcolor={blue!80!black}
}
\usepackage{bm}
\usepackage{epstopdf}
\usepackage{soul}
\usepackage{epsfig}
\usepackage{bbold}
\usepackage{braket}
\usepackage{physics}
\usepackage{gensymb}
\usepackage{amsmath,amssymb}

\begin{document}

\title{Spin angular momentum and optical chirality of Poincar\'e vector vortex beams}% Force line breaks with \\
\author{Kayn A. Forbes}
\email{K.Forbes@uea.ac.uk}

\affiliation{School of Chemistry, University of East Anglia, Norwich Research Park, Norwich NR4 7TJ, United Kingdom}

\begin{abstract}
 The optical chirality and spin angular momentum of structured scalar vortex beams has been intensively studied in recent years. The pseudoscalar topological charge $\ell$ of these beams is responsible for their unique properties. Constructed from a superposition of scalar vortex beams with topological charges $\ell_\text{A}$ and $\ell_\text{B}$, cylindrical vector vortex beams are higher-order Poincar\'e  modes which possess a spatially inhomogeneous polarization distribution. Here we highlight the highly tailorable and exotic spatial distributions of the optical spin and chirality densities of these higher-order structured beams under both paraxial (weak focusing) and non-paraxial (tight focusing) conditions. Our analytical theory can yield the spin angular momentum and optical chirality of each point on any higher-order or hybrid-order Poincar\'e sphere. It is shown that the tunable Pancharatnam topological charge $\ell_{\text{P}} = (\ell_\text{A} + \ell_\text{B})/2$ and polarization index $m = (\ell_\text{B} -\ell_\text{A})/2$ of the vector vortex beam plays a decisive role in customizing their spin and chirality spatial distributions. We also provide the correct analytical equations to describe a focused, non-paraxial scalar Bessel beam.  
\end{abstract}

\maketitle

\section{Introduction}

Optical beams can carry energy, linear momentum,  angular momentum (both spin and orbital), and optical chirality. These  properties of the electromagnetic fields are conserved for beams in free space. They manifest in light-matter interactions in a number of ways: for example, the angular momentum can create torques on particles, while the optical chirality is partly responsible for chiral light-matter interactions and optical activity. The perceived canonical directions and magnitudes of these optical properties have been ingrained through the ubiquitous plane-wave electromagnetic field description of light studied in textbooks. However, light in general is significantly more structured in its degrees of freedom (amplitude, phase, polarization) than a plane wave. Optical fields with more complex spatial distributions of their degrees of freedom readily carry extraordinary optical properties when compared to a plane wave. A well-known type is the optical vortex beam with its azimuthal phase $\text{e}^{i\ell\phi}$, where $\ell \in \mathbb{Z}$ is the topological charge, leading to an orbital angular momentum (OAM) of $\ell \hbar$ per photon in the drection of propagation \cite{shen2019optical}.

Structured light \cite{andrews2011structured, forbes2019structured, forbes2021structured, he2022towards} refers to our ability to tailor the amplitude, phase, and polarization degrees of freedom in both space and time, leading to a remarkable diversity of laser beam structure. One important type of structured light are vector beams. In comparison to scalar beams which possess an homogeneous spatial distribution of polarization state, vector beams have spatially inhomogenous polarization states. More generally, vector beams are referred to as vector vortex beams (VVBs) \cite{rosales2018review, arora2020detection}. VVBs are beams which possess both inhomogeneous polarization structure and azimuthal phase structure. VVBs can be viewed as the superposition of two scalar orthogonally polarized spatial modes (labelled A and B) which individually carry optical OAM through the azimuthal phase factor $\text{e}^{i\ell_{\text{A(B)}}\phi}$. VVBs are also known as higher-order Poincar\'e  (HOP) \cite{milione2011higher} or hybrid-order Poincar\'e  (HyOP) \cite{yi2015hybrid, arora2020detection, ling2016characterization, arora2020hybrid} beams, the former carry the same magnitude of OAM but of opposite sign $\ell_{\text{A}} = -\ell_{\text{B}}$: cylindrical vector beams (CVBs). The latter carry arbitrary OAM: cylindrical vector vortex beams (CVVBs). The controllable combination of the different degrees of freedom of these significantly structured light beams means they under intense study due to their applicability in areas such as optical manipulation, optical communications, quantum information, laser machining, and enhanced imaging to name a few  \cite{naidoo2016controlled, forbes2019quantum, wang2020vectorial, he2022towards, nape2022revealing, nape2023quantum,bliokh2019geometric, wang2021polarization}.  

In this work we provide a general, simple analytical analysis of the  optical spin and chirality properties of CVBs and CVVBs valid for both paraxial (weakly focused) and nonparaxial (tightly focused) light. Our theoretical description uses orthogonally polarized Bessel beam modes expanded in a smallness parameter to quantify the degree of focusing. Our theory allows the extraction of the spin and chirality of each point on any higher-order or hybrid-order Poincar\'e sphere. We also provide the correct electromagnetic fields to describe a focused scalar Bessel beam and comment on the previous forms found in the literature.     

\section{Theoretical description of Bessel Vector Vortex Beams}

Vector beams can be written as a superposition of two co-propagating (we assume along $z$) orthogonally polarized scalar beams \cite{andrews2012angular, rosales2018review}. The amplitude distribution is free to be chosen as any solution to the wave equation: Laguerre-Gaussian, Hermite-Gaussian, Bessel, Bessel-Gauss, etc. In this work we use pure Bessel modes due to their analytical simplicity and the fact they are solutions to both the paraxial and nonparaxial wave equations. The electric field for a monochromatic scalar Bessel beam up to second-order in the paraxial parameter $k_t/k_z$ is 

\begin{widetext}
\begin{align}
\mathbf{E} &= \Bigl[ J_{|{\ell}|}\text{e}^{i\ell\phi}(\alpha\mathbf{\hat{x}}+\beta\mathbf{\hat{y}}) \nonumber + \mathbf{\hat{z}}\frac{ik_t}{2k_z}\bigl((\alpha\pm i \beta)J_{|\ell|-1}\text{e}^{i(\ell\mp1)\phi} +(\pm i \beta-\alpha)J_{|\ell|+1}\text{e}^{i(\ell\pm1)\phi}\bigl) \nonumber + \mathbf{\hat{x}}\frac{k_t^2}{4k^2}\Bigl(2\alpha J_{|\ell|}\text{e}^{i\ell\phi} \nonumber \\
&+ J_{|\ell|-2}(\alpha \pm i \beta)\text{e}^{i(\ell\mp2)\phi} + J_{|\ell|+2}(\alpha \mp i \beta)\text{e}^{i(\ell\pm2)\phi}\Bigr) + \mathbf{\hat{y}}\frac{k_t^2}{4k^2}\Bigl(2\beta J_{|\ell|}\text{e}^{i\ell\phi} \nonumber + J_{|\ell|-2}(\pm i \alpha - \beta)\text{e}^{i(\ell\mp2)\phi} \nonumber \\
&+ J_{|\ell|+2}(\mp i \alpha -\beta)\text{e}^{i(\ell\pm2)\phi}\Bigr)\Bigr]\text{e}^{ik_z z}, 
\label{eq:1}
\end{align}
\end{widetext}

where $J_{|\ell|}[k_t r]$ is a Bessel function of the first-kind of order $|\ell|$ and argument $k_t r$ (the argument is suppressed in Eq. \eqref{eq:1} and throughout the manuscript for notational brevity, further we subsume units of electric field into the Bessel function); $\ell \in \mathbb{Z}$ is the topological charge, $\ell >0$ left-handed helical wavefronts, $\ell < 0$  right-handed helical wavefronts; $\phi$ is the azimuthal angle; $\alpha$ and $\beta$ are the Jones vector coefficients; $k_z=\sqrt{k^2-k_t^2}$ is the longitudinal wavenumber and $k_t=\sqrt{k_x^2+k_y^2}$ the transverse wavenumber. The  derivation of Eq. \eqref{eq:1} (and Eq. \eqref{eq:2}) is accomplished using the Supplementary Information of \cite{forbes2023customized} in conjunction with Appendix III of this paper. The rule determining which sign to take for the $\pm$ and $\mp$ parts in Eq.~\eqref{eq:1} (and \eqref{eq:2}) is that if the topological charge of the mode is $\ell>0$ the upper-sign is taken; if $\ell<0$ the lower sign is taken. Note that in this work we explicitly use the circular polarization basis such that $\alpha = 1/\sqrt{2}$ and $\beta = i\sigma /\sqrt{2}$, where the helicity is $\sigma = \pm 1$, the positive denoting left-handed circular polarization, the negative sign right-handed.

In language first introduced by Lax et al. \cite{lax1975maxwell},  Eq. \eqref{eq:1} contains the zeroth-order transverse $\text{T}_0$ (with respect to the smallness parameter $k_t/k_z)$, first-order longitudinal $\text{L}_1$, and second-order transverse field components $\text{T}_2$. The zeroth-order term in Eq. \eqref{eq:1} is the dominating term for a paraxial (well-collimated) Bessel beam; as a Bessel beam is spatially confined (focused) the ratio of $k_t/k_z$ becomes larger whereupon the higher-order field components, first-order longitudinal and second-order transverse, become significant enough in magnitude compared to the zeroth-order fields to yield physically observable effects \cite{forbes2021relevance}. This transition from paraxial optics to nonparaxial optics leads to the rich behaviour of spatially confined electromagnetic fields in nano-optics. Most clear to see is that the 2D ($x,y$) polarized paraxial beam becomes 3D ($x,y,z$) polarized \cite{alonso2023geometric}. It is important to note that all of the higher-order fields are ever-present even in a well-collimated, paraxial beam of light, however their magnitude is essentially zero with respect to the dominating zeorth-order transverse field under weak focusing conditions.  

Maxwell's equations in free-space are dual symmetric, however due to the electric-bias nature of most dielectric materials, the magnetic field is little studied compared to the electric field. For example, the magnetic contributions to the energy, momentum, and angular momentum of the field. However, in this work we look at the optical chirality, a conserved property of the free-field which is the inner product of the electric and magnetic field. We therefore require the corresponding magnetic field of a Bessel beam: 

\begin{widetext}
\begin{align}
\mathbf{B} &= \Bigl[J_{|{\ell}|}\text{e}^{i\ell\phi} \frac{k_z}{k}(\alpha\mathbf{\hat{y}}-\beta\mathbf{\hat{x}}) + \mathbf{\hat{z}}\frac{ik_t}{2k}\bigl((\pm i\alpha - \beta)J_{|\ell|-1}\text{e}^{i(\ell\mp1)\phi} +(\pm i \alpha + \beta)J_{|\ell|+1}\text{e}^{i(\ell\pm1)\phi}\bigl) + \mathbf{\hat{x}}\frac{k_t^2}{4kk_z}\Bigl(-2\beta J_{|\ell|}\text{e}^{i\ell\phi} \nonumber \\
&+ J_{|\ell|-2}(\pm i \alpha - \beta)\text{e}^{i(\ell\mp2)\phi} + J_{|\ell|+2}(\mp i \alpha - \beta)\text{e}^{i(\ell\pm2)\phi}\Bigr) + \mathbf{\hat{y}}\frac{k_t^2}{4kk_z}\Bigl(2\alpha J_{|\ell|}\text{e}^{i\ell\phi} + J_{|\ell|-2}(\mp i \beta - \alpha)\text{e}^{i(\ell\mp2)\phi} \nonumber \\ 
&+ J_{|\ell|+2}(\pm i \beta -\alpha)\text{e}^{i(\ell\pm2)\phi}\Bigr)\Bigr]\frac{1}{c}\text{e}^{ik_z z}.
\label{eq:2}
\end{align}
\end{widetext}

The analytical electromagnetic fields Eqs. \eqref{eq:1} and \eqref{eq:2} containing field components up to second-order in the smallness parameter are used in this manuscript to describe CVBs and CVVBs. An alternative and widely used approach currently is to use numerical integration methods based on Richards-Wolf diffraction theory \cite{novotny2012principles} or the Ignatovsky \cite{peatross2017vector} model to describe the electromagnetic fields of a focused beam. These numerical techniques have been widely used to study the optical properties of vector beams \cite{zhu2022evolving, zhu2014spin, li2021spin, zhang2022ultrafast, shi2018structured, meng2019angular, huang2011vector, zhu2023generation}. A detailed comparison of the differing methods can be found in Ref. \cite{peatross2017vector}, but essentially the analytical methods we favour in this work lead to simple analytical results and a deep insight into the novel contributions from specific higher-order field components to properties of electromagnetic fields.

\section{Spin Angular Momentum}

The spin angular momentum of light can be both longitudinal and transverse with respect to the direction of propagation \cite{andrews2012angular,bliokh2015transverse,aiello2015transverse}. Longitudinal spin angular momentum is much more familiar, and the spin of $\sigma \hbar \mathbf{\hat{z}}$ per photon for a $z$-propagating circularly polarized plane wave with helicity $\sigma = \pm 1$ is a well-known result. More extraordinary is the transverse spin of light, underpinning chiral quantum optics \cite{lodahl2017chiral}, spin-momentum locking \cite{bliokh2015spin}, and the quantum Hall effect of light \cite{bliokh2015quantum}. The cycle-averaged (electric) spin momentum density for a monochromatic beam is calculated using \cite{bliokh2013dual}

\begin{align}
\mathbf{s_E} = \frac{\epsilon_0}{2} \text{Im}\mathbf{E}^*\times\mathbf{E}. 
\label{eq:3}
\end{align}

The longitudinal spin angular momentum is generated by the cross product between the transverse $(x,y)$ polarized fields. The transverse spin of light manifests through the cross product of the transverse field with the $z$-polarized longitudinal component. In this work we calculate the optical properties of VVBs up to second-order in the smallness parameter: for the spin angular momentum this therefore includes the $\text{T}_0 \times \text{T}_0$, $\text{T}_0 \times \text{L}_1$, and $\text{T}_0 \times \text{T}_2$ contributions, and neglects the extremely small higher-order contributions.

\subsection{$ \ell_\text{A}^{\text{R}} + \ell_\text{B}^{\text{L}} $ vector beams}

The first type of vector beam we look at consists of a superposition of a left circularly polarized beam A and a right-circularly polarized beam B where each beam has the opposite signed topological charges to their polarization helicity, i.e. $ \text{sgn} \ell_\text{A} =  \text{sgn} \sigma_\text{B}$ and $ \text{sgn} \ell_\text{B} =  \text{sgn} \sigma_\text{A}$. Thus A and B also have opposite signed topological charges with respect to one another $ \text{sgn} \ell_\text{A} = - \text{sgn} \ell_\text{B}$, i.e. $ \ell_\text{A}^{\text{R}} + \ell_\text{B}^{\text{L}} $, where the superscript labels R and L correspond to right (left) wavefront handedness. Note there is also the case of $ \text{sgn} \ell_\text{A} = - \text{sgn} \ell_\text{B}$, but where $ \text{sgn} \ell_\text{A} =  \text{sgn} \sigma_\text{A}$ and $ \text{sgn} \ell_\text{B} =  \text{sgn} \sigma_\text{B}$, i.e. $ \ell_\text{A}^{\text{L}} + \ell_\text{B}^{\text{R}} $: the results for these beams can be found in the Appendix I.

As mentioned in the Introduction, vector beams in general are referred to as HyOP beams or CVVBs. The so-called HOP beams, or CVBs, are a subset of CVVBs/HyOP beams where $\ell_\text{A} = - \ell_\text{B}$, i.e. the topological charges have opposite sign and equal magnitude. The electric field of a CVVB  produced by a superposition of two Bessel beams can therefore be extracted from Eq. \eqref{eq:1} and is given explicitly as:

\begin{widetext}
\begin{align}
\mathbf{E} &= \Bigl(J_{|\ell_\text{A}|}\text{e}^{-i(|\ell_\text{A}|\phi+\theta)}\sin\chi \begin{bmatrix}
1\\
i
\end{bmatrix} 
+
J_{|\ell_\text{B}|}\text{e}^{i(|\ell_\text{B}|\phi+\theta)} \cos\chi \begin{bmatrix}
1\\
-i
\end{bmatrix} + \mathbf{\hat{z}}\frac{ik_t}{k_z}\bigl(J_{|\ell_\text{A}|-1}\text{e}^{i[(1-|\ell_\text{A}|)\phi-\theta]}\sin\chi+J_{|\ell_\text{B}|-1}\text{e}^{i[(|\ell_\text{B}|-1)\phi+\theta]}\cos\chi\bigl) \nonumber\\ 
&+ \frac{k_t^2}{2k^2}\big[\mathbf{\hat{x}}(J_{|\ell_\text{A}|}\text{e}^{-i(|\ell_\text{A}|\phi+\theta)}\sin\chi + J_{|\ell_\text{B}|}\text{e}^{i(|\ell_\text{B}|\phi+\theta)}\cos\chi + J_{|\ell_\text{A}|-2}\text{e}^{i[(2-|\ell_\text{A}|)\phi-\theta]}\sin\chi + J_{|\ell_\text{B}|-2}\text{e}^{i[(|\ell_\text{B}|-2)\phi+\theta]}\cos\chi) \nonumber \\
&+i\mathbf{\hat{y}}(J_{|\ell_\text{A}|}\text{e}^{-i(|\ell_\text{A}|\phi+\theta)}\sin\chi - J_{|\ell_\text{B}|}\text{e}^{i(|\ell_\text{B}|\phi+\theta)}\cos\chi - J_{|\ell_\text{A}|-2}\text{e}^{i[(2-|\ell_\text{A}|)\phi-\theta]}\sin\chi + J_{|\ell_\text{B}|-2}\text{e}^{i[(|\ell_\text{B}|-2)\phi+\theta]}\cos\chi)\bigr]\Bigr)\text{e}^{ik_z z}.
\label{eq:4}
\end{align}
\end{widetext}

\begin{figure*}
    \includegraphics[]{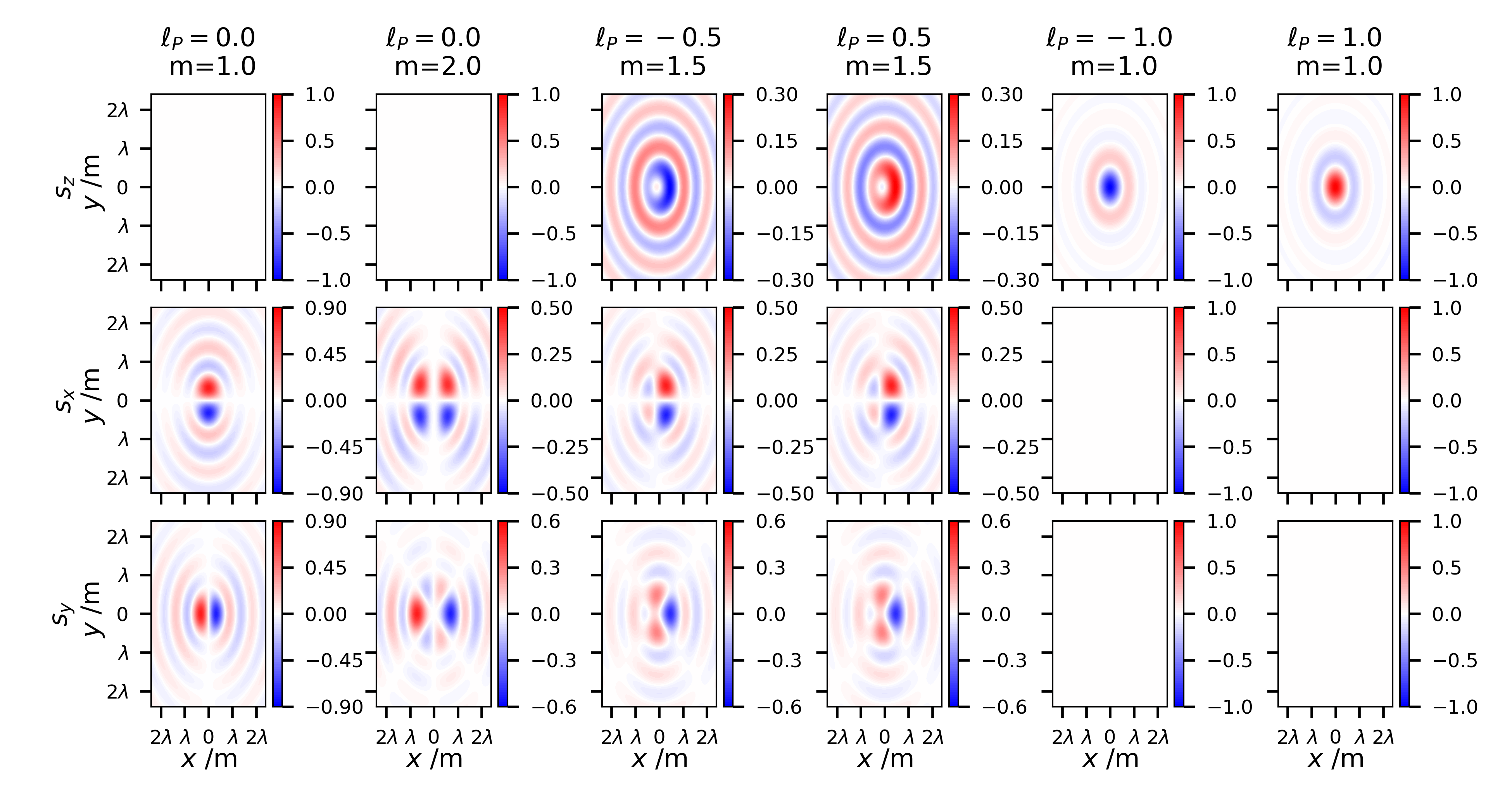}
    \caption{Spin angular momentum density spatial distributions of $ \ell_\text{A}^{\text{R}} + \ell_\text{B}^{\text{L}} $ vector beams: Top row: $z$-component Eq.~\eqref{eq:5}; Middle: $x$-component Eq.~\eqref{eq:6}; Bottom: $y$-component Eq.~\eqref{eq:7}. The simulated parameters throughout this paper are as follows: $k_t/k_z = 0.6315$, $\lambda = 729 \text{nm}$ i.e. tight focus; The colour scale indicates the intensity in arbitrary units; $\theta = 0$ and $\chi = \pi/4$. }
    \label{fig:1}
\end{figure*}

The introduction of the phase $\text{e}^{\pm i \theta}$, where $\theta \in [0,\pi]$ describes the longitude on the Poincar\'e sphere. \cite{galvez2012poincare, andrews2012angular}. For example, for $|\ell_{(\text{A/B})}| = 1$, $\theta = 0$ corresponds to first-order radially polarized beams, whereas $\theta = \pi/2$ produce first-order azimuthally polarized beams. The angle $\chi \in [0,\pi/2]$ describes the lattitude on the Poincar\'e sphere. For a given arbitrary Poincar\'e sphere, the general pure state of 2D polarization (that of the zeroth-order transverse field) can be described by the angles $\chi$ and $\theta$: the former tells us the shape of the polarization ellipse, the latter its orientation. 

Using the electric field Eq. \eqref{eq:4} in Eq. \eqref{eq:3} gives for the $z$ component (up to second-order in the smallness parameter) of the spin angular momentum density

\begin{align}
{s^E_z} &= (J_{|\ell_\text{A}|}^2\sin^2\chi-J_{|\ell_\text{B}|}^2\cos^2\chi) (1+\frac{2k_t^2}{k^2}) \nonumber \\
&+\frac{k_t^2}{k^2}(J_{|\ell_\text{A}|}J_{|\ell_\text{B}|-2}-J_{|\ell_\text{B}|}J_{|\ell_\text{A}|-2}) \nonumber \\ &\times\cos[(|\ell_\text{A}|+|\ell_\text{B}|-2)\phi+2\theta]\sin2\chi, 
\label{eq:5}
\end{align}

for the $x$ component,

\begin{align}
{s^E_x} &= \frac{k_t}{k_z}\bigl[2(J_{|\ell_\text{A}|}J_{|\ell_\text{A}|-1}\sin^2\chi + J_{|\ell_\text{B}|}J_{|\ell_\text{B}|-1}\cos^2\chi)\sin\phi \nonumber \\
&+(J_{|\ell_\text{A}|}J_{|\ell_\text{B}|-1} +J_{|\ell_\text{B}|}J_{|\ell_\text{A}|-1}) \nonumber \\ 
& \times \sin \bigl((|\ell_\text{A}|+|\ell_\text{B}|-1)\phi + 2\theta\bigr)\sin2\chi\bigr],
\label{eq:6}
\end{align}

and the $y$ component, 

\begin{align}
{s^E_y} &= -\frac{k_t}{k_z}\bigl[2(J_{|\ell_\text{A}|}J_{|\ell_\text{A}|-1}\sin^2\chi + J_{|\ell_\text{B}|}J_{|\ell_\text{B}|-1}\cos^2\chi)\cos\phi \nonumber \\
&+(J_{|\ell_\text{A}|}J_{|\ell_\text{B}|-1} +J_{|\ell_\text{B}|}J_{|\ell_\text{A}|-1}) \nonumber \\ & \times \cos\bigl((|\ell_\text{A}|+|\ell_\text{B}|-1)\phi + 2\theta\bigr)\sin2\chi].
\label{eq:7}
\end{align}

Firstly it is clear that the amplitude of A and B have significant influence upon the properties of the field. However, it is more interesting, and given how these beams are generated experimentally \cite{naidoo2016controlled, cardano2015spin, sroor2020high, zhan2009cylindrical, liu2021broadband, feng2023generation, liu2017generation, ling2016characterization, galvez2012poincare, alpmann2017dynamic}, more relevant to assume equal amplitudes and vary other parameters of the beam. When $\chi = 0$ or $\pi/2$, Eqs.~\eqref{eq:5}-\eqref{eq:7} describe the spin of a scalar right-circularly polarized with $\ell>0$ and left-circularly polarized with $\ell<0$ Bessel beam, respectively.

In the case of weak focusing, $k_t << k_z$ and $k \approx k_z$ which means that the contributions from the zeroth-order transverse field components are far larger than the higher-order first-order longitudinal and second-order transverse components, i.e. paraxial optics. Under such conditions the transverse spin is extremely weak, and the longitudinal spin is simply proportional to the difference in amplitude between beam A and beam B. The optical chirality (which we look at in the next section) is likewise proportional to the difference between the amplitudes of beam A and beam B, and in paraxial optics there is a simple proportionality between the polarization ellipticity, chirality, and spin, often measured by the fourth Stokes parameter $S_3$.  

\begin{figure*}
    \includegraphics[]{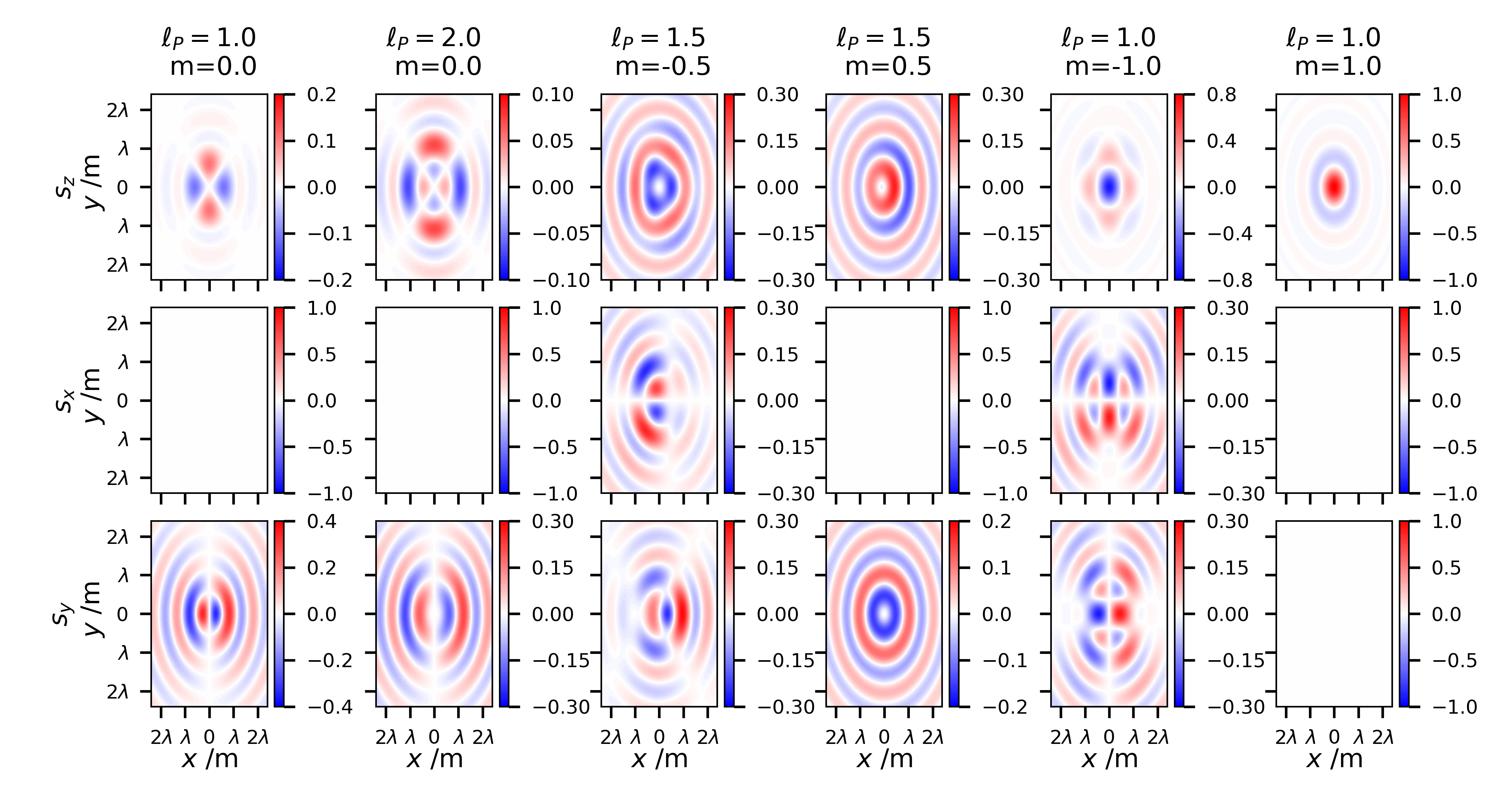}
    \caption{Spin angular momentum density spatial distributions of $ \ell_\text{A}^{\text{L}} + \ell_\text{B}^{\text{L}} $ vector beams: Top row: $z$-component Eq.~\eqref{eq:9}; Middle: $x$-component Eq.~\eqref{eq:10}; Bottom: $y$-component Eq.~\eqref{eq:11}.}
    \label{fig:2}
\end{figure*}

We can use the Pancharatnam topological charge $\ell_{\text{P}} = (\ell_{\text{A}}+\ell_{\text{B}})/2$ \cite{niv2006manipulation, cohen2019geometric, li2023generation} and polarization order $m =(\ell_{\text{B}}-\ell_{\text{A}})/2$ to characterize VVBs. Clearly for CVBs (or HOPs) $\ell_{\text{P}} = 0$. For $ \text{sgn} \ell_\text{A} = - \text{sgn} \ell_\text{B}$ beams, where $\text{sgn} \ell_\text{A} = \text{sgn} \sigma_\text{B}$, $\text{sgn} \ell_\text{B} = \text{sgn} \sigma_\text{A}$, then  $\ell_{\text{P}} = (|\ell_{\text{B}}|-|\ell_{\text{A}}|)/2$ and $m =(|\ell_{\text{A}}|+|\ell_{\text{B}}|)/2$. The spatial distributions of Eqs.~\eqref{eq:5}-\eqref{eq:7} are given in Fig. \ref{fig:1} for a varying range of $\ell_\text{P}$ and $m$. Furthermore, throughout this work, we highlight in the Figures the spatial distributions at the specific point $\chi=\pi/4$ (on the equator) and $\theta=0$ on any arbitrary higher-order/hybrid Poincar\'e sphere, however it is important to note the equations are kept completely general and can evaluate each point on any Poincar\'e sphere. Fig. \ref{eq:1} shows that there is no $z$-component of spin for $\ell_{\text{P}} = 0$ CVBs, however there are, in general, transverse spin components in both $x$ and $y$. Unlike $s_z$, the transverse spin is independent of the sign (or handedness) of $\ell_{\text{P}}$, but is clearly very sensitive to both the magnitude of $\ell_{\text{P}}$ and the polarization index $m$. There is a particularly strong component of transverse spin in the case of $\ell_{\text{P}} = 0$ and $m = 1$, which is not surprising given this mode represents the first-order radially polarized vector beam, well-known for its strong longitudinal electric field component and resultant transverse spin \cite{neugebauer2015measuring}. 

\subsection{$\ell_\text{A}^{\text{L}} + \ell_\text{B}^{\text{L}}$ vector beams}

We now look at CVVBs consisting of a left- and right-handed circularly polarized superposition where each beam has the same signed topological charges, i.e. $ \text{sgn} \ell_\text{A} =  \text{sgn} \ell_\text{B}$. It is important to note that here we look at the case of  $\ell_\text{A}$ and $\ell_\text{B}$ both being positive, i.e. left-handed $\ell_\text{A}^{\text{L}} + \ell_\text{B}^{\text{L}}$: the results are different if both $\ell_\text{A}$ and $\ell_\text{B}$ are right-handed, i.e. $\ell_\text{A}^{\text{R}} + \ell_\text{B}^{\text{R}}$ (see Appendix II). The electric field of these CVVBs produced by a superposition of two Bessel beams can therefore be extracted from Eq. \eqref{eq:1} and is given explicitly as:

\begin{widetext}
\begin{align}
\mathbf{E} &= \Bigl(J_{|\ell_\text{A}|}\text{e}^{i(|\ell_\text{A}|\phi-\theta)} \sin\chi \begin{bmatrix}
1\\
i
\end{bmatrix} 
+
J_{|\ell_\text{B}|}\text{e}^{i(|\ell_\text{B}|\phi+\theta)} \cos\chi \begin{bmatrix}
1\\
-i
\end{bmatrix} + \mathbf{\hat{z}}\frac{ik_t}{k_z}\bigl(J_{|\ell_\text{B}|-1}\text{e}^{i[(|\ell_\text{B}|-1)\phi+\theta]} \cos\chi -J_{|\ell_\text{A}|+1}\text{e}^{i[(|\ell_\text{A}|+1)\phi-\theta]} \sin\chi \bigl) \nonumber \\ \nonumber
&+ \frac{k_t^2}{2k^2}\big[\mathbf{\hat{x}}(J_{|\ell_\text{A}|}\text{e}^{i(|\ell_\text{A}|\phi-\theta)} \sin\chi + J_{|\ell_\text{B}|}\text{e}^{i(|\ell_\text{B}|\phi+\theta)} \cos\chi + J_{|\ell_\text{A}|+2}\text{e}^{i[(|\ell_\text{A}|+2)\phi-\theta]} \sin\chi + J_{|\ell_\text{B}|-2}\text{e}^{i[(|\ell_\text{B}|-2)\phi+\theta]} \cos\chi) \\ 
&+i\mathbf{\hat{y}}(J_{|\ell_\text{A}|}\text{e}^{i(|\ell_\text{A}|\phi-\theta)} \sin\chi - J_{|\ell_\text{B}|}\text{e}^{i(|\ell_\text{B}|\phi+\theta)} \cos\chi - J_{|\ell_\text{A}|+2}\text{e}^{i[(|\ell_\text{A}|+2)\phi-\theta]}\sin\chi + J_{|\ell_\text{B}|-2}\text{e}^{i[(|\ell_\text{B}|-2)\phi+\theta]}\cos\chi)\bigr]\Bigr)\text{e}^{ik_z z}.
\label{eq:8}
\end{align}
\end{widetext}

Inserting Eq. \eqref{eq:8} into Eq. \eqref{eq:3} gives the $z$ component of the spin angular momentum density as,

\begin{align}
{s^E_z} &= (J_{|\ell_\text{A}|}^2\sin^2\chi-J_{|\ell_\text{B}|}^2\cos^2\chi) (1+\frac{2k_t^2}{k^2}) \nonumber \\ \nonumber
&+\frac{k_t^2}{k^2}(J_{|\ell_\text{A}|}J_{|\ell_\text{B}|-2}-J_{|\ell_\text{B}|}J_{|\ell_\text{A}|+2}) \\ 
& \times \cos[(|\ell_\text{A}|-|\ell_\text{B}|+2)\phi-2\theta]\sin2\chi, 
\label{eq:9}
\end{align}

the $x$ component as,

\begin{align}
{s^E_x} &= \frac{k_t}{k_z}\bigl[2(J_{|\ell_\text{B}|}J_{|\ell_\text{B}|-1}\cos^2\chi -J_{|\ell_\text{A}|}J_{|\ell_\text{A}|+1}\sin^2\chi)\sin\phi \nonumber\\ 
&+(J_{|\ell_\text{B}|}J_{|\ell_\text{A}|+1}-J_{|\ell_\text{A}|}J_{|\ell_\text{B}|-1}) \nonumber \\ &\times \sin\bigl((|\ell_\text{A}|-|\ell_\text{B}|+1)\phi - 2\theta\bigr)\sin2\chi],
\label{eq:10}
\end{align}

and the $y$ component, 

\begin{align}
{s^E_y} &= -\frac{k_t}{k_z}\bigl[2(J_{|\ell_\text{B}|}J_{|\ell_\text{B}|-1}\cos^2\chi -J_{|\ell_\text{A}|}J_{|\ell_\text{A}|+1}\sin^2\chi)\cos\phi \nonumber\\ 
&+(J_{|\ell_\text{A}|}J_{|\ell_\text{B}|-1}-J_{|\ell_\text{B}|}J_{|\ell_\text{A}|+1}) \nonumber \\
&\times \cos\bigl((|\ell_\text{A}|-|\ell_\text{B}|+1)\phi - 2\theta\bigr)\sin2\chi].
\label{eq:11}
\end{align}

For $ \text{sgn} \ell_\text{A} = \text{sgn} \ell_\text{B}$ beams, where $\text{sgn} \ell_\text{A} = \text{sgn} \sigma_\text{A}$, $\text{sgn} \ell_\text{B} = \text{sgn} \sigma_\text{A}$, then  $\ell_{\text{P}} = (|\ell_{\text{A}}| + |\ell_{\text{B}}|)/2$ and $m =(|\ell_{\text{B}}|-|\ell_{\text{A}}|)/2$. The spatial distributions of Eqs.~\eqref{eq:9}-\eqref{eq:11} are given in Fig. \ref{fig:2} for a varying range of $\ell_\text{P}$ and $m$. When $\chi = 0$ or $\pi/2$, Eqs.~\eqref{eq:9}-\eqref{eq:11} describe the spin of a scalar right-circularly polarized with $\ell>0$ and left-circularly polarized with $\ell>0$ Bessel beam, respectively. In contrast to Figure \ref{fig:1}, this class of CVVB can carry transverse spin in a single direction and not the other, as well as circularly-symmetric spatial distributions in a single direction (i.e. $s_y$ for $\ell_\text{P} = 1.5$ and $m =0.5$). Similar to Figure \ref{fig:1} both $\ell_\text{P}$ and $m$ strongly influence the spatial distribution and sign of the spin angular momentum density. 

\begin{figure*}
    \includegraphics[]{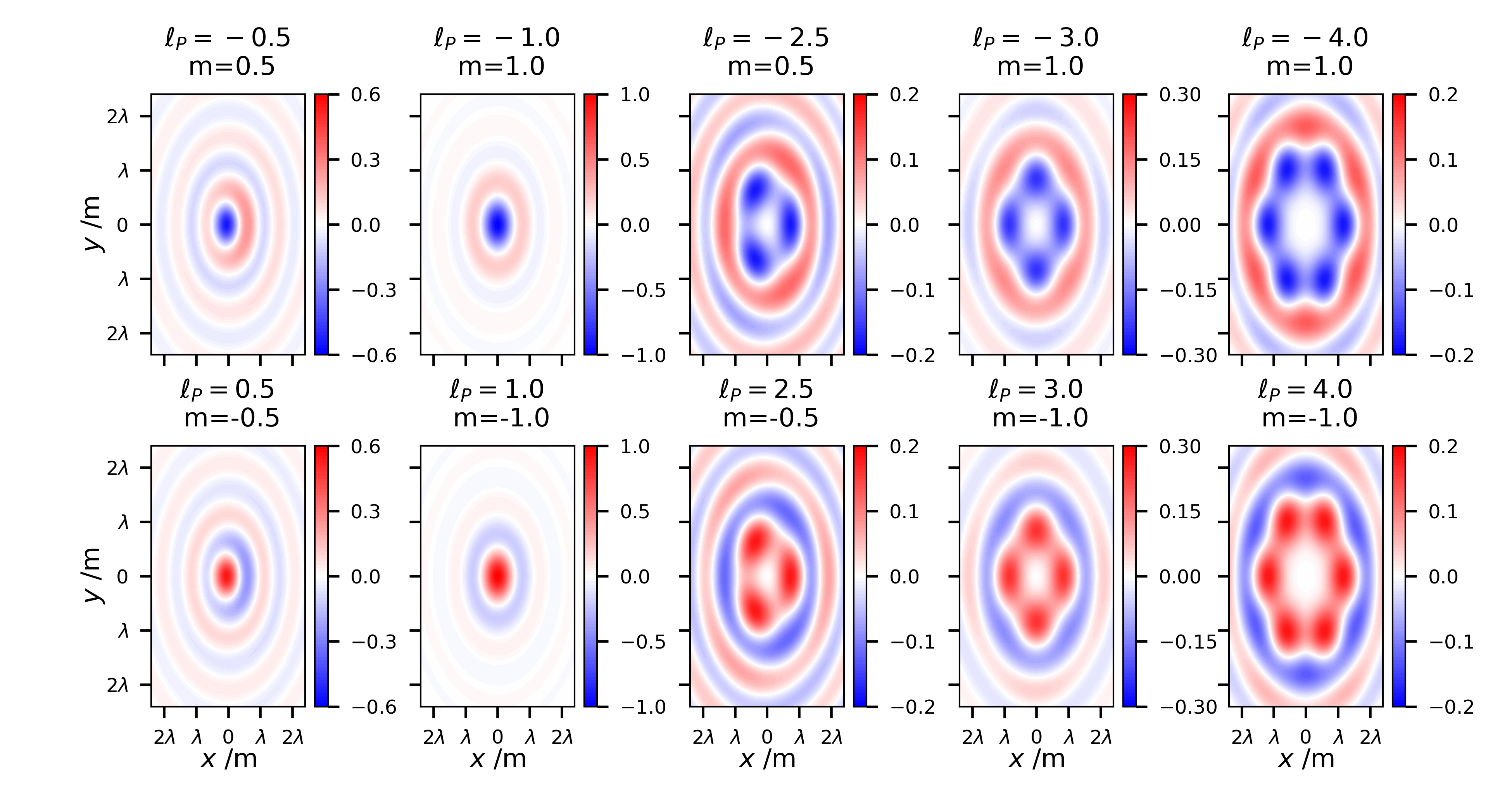}
    \caption{Optical chirality density spatial distributions of $ \ell_\text{A}^{\text{R}} + \ell_\text{B}^{\text{L}} $ vector beams Eq~.\eqref{eq:14}: Top row: $\ell_{\text{A}} > \ell_{\text{B}}$; Bottom: $\ell_{\text{A}} < \ell_{\text{B}}$, i.e. negative $\ell_{\text{P}}$ of Top row.}
    \label{fig:3}
\end{figure*}

\section{Optical chirality}

The optical chirality density of a monochromatic beam is \cite{bliokh2013dual} 

\begin{align}
C = - \frac{\epsilon_0 \omega}{2}\text{Im}\mathbf{E}^*\cdot\mathbf{B}. 
\label{eq:12}
\end{align}

For monochromatic beams, the pseudoscalar optical chirality is directly proportional to optical helicity \cite{cameron2012optical, mackinnon2019differences}. The optical chirality is the electromagnetic property which couples to the electric-dipole (E1) magnetic-dipole (M1) interference term in the linear absorption and scattering of light by chiral materials. As such, it does not describe all chiral light-matter interactions (e.g. nonlinear interactions or those due to electric-dipole electric quadrupole (E2) interference E1E2). Optical chirality Eq. \eqref{eq:12} should not be confused with the more general geometrical meaning of chirality. For example, linearly polarized, collimated scalar optical vortex modes possess a geometrical chirality due to their twisted wavefront, being left-handed for $\ell >0$ and right-handed for $\ell<0$. However, they possess zero optical chirality. Only under non-paraxial, tight focusing conditions does the chirality associated with the pseudoscalar $\ell$ contribute to the optical chirality \cite{forbes2021measures, forbes2021optical, green2023optical, forbes2023customized}. As with the spin angular momentum, we will calculate the optical chirality up to second-order in the smallness parameter. Therefore we include the $\text{T}_0 \cdot \text{T}_0$, $\text{L}_1 \cdot \text{L}_1$, and $\text{T}_0 \cdot \text{T}_2$ contributions, and neglect the extremely small higher-order contributions. \\

\subsection{$\ell_\text{A}^{\text{R}} + \ell_\text{B}^{\text{L}}$ vector beams}

We already have the electric field which relates to CVVBs Eq. \eqref{eq:4} of type $\ell_\text{A}^{\text{R}} + \ell_\text{B}^{\text{L}}$. In order to calculate the optical chirality density using Eq. \eqref{eq:12} we require the corresponding magnetic field. This magnetic field is extracted from Eq. \eqref{eq:2} as

\begin{widetext}
\begin{align}
\mathbf{B} &=\Bigl(J_{|\ell_\text{A}|}\text{e}^{-i(|\ell_\text{A}|\phi+\theta)} \frac{k_z}{k} \sin\chi \begin{bmatrix}
-i\\
1
\end{bmatrix} +
J_{|\ell_\text{B}|}\text{e}^{i(|\ell_\text{B}|\phi+\theta)} \frac{k_z}{k} \cos\chi \begin{bmatrix}
i\\ \nonumber
1
\end{bmatrix}  + \mathbf{\hat{z}}\frac{k_t}{k}\bigl(J_{|\ell_\text{A}|-1}\text{e}^{i[(1-|\ell_\text{A}|)\phi-\theta]}\sin\chi \nonumber \\
&-J_{|\ell_\text{B}|-1}\text{e}^{i[(|\ell_\text{B}|-1)\phi+\theta]}\cos\chi\bigl) + \frac{k_t^2}{2kk_z}\big[i\mathbf{\hat{x}}(-J_{|\ell_\text{A}|}\text{e}^{-i(|\ell_\text{A}|\phi+\theta)}\sin\chi + J_{|\ell_\text{B}|}\text{e}^{i(|\ell_\text{B}|\phi+\theta)}\cos\chi \nonumber \\ \nonumber &- J_{|\ell_\text{A}|-2}\text{e}^{i[(2-|\ell_\text{A}|)\phi-\theta]}\sin\chi + J_{|\ell_\text{B}|-2}\text{e}^{i[(|\ell_\text{B}|-2)\phi+\theta]}\cos\chi) +\mathbf{\hat{y}}(J_{|\ell_\text{A}|}\text{e}^{-i(|\ell_\text{A}|\phi+\theta)}\sin\chi + J_{|\ell_\text{B}|}\text{e}^{i(|\ell_\text{B}|\phi+\theta)}\cos\chi \\  &- J_{|\ell_\text{A}|-2}\text{e}^{i[(2-|\ell_\text{A}|)\phi-\theta]}\sin\chi - J_{|\ell_\text{B}|-2}\text{e}^{i[(|\ell_\text{B}|-2)\phi+\theta]}\cos\chi)\bigr]\Bigr)\frac{1}{c}\text{e}^{ik_z z}.
\label{eq:13}
\end{align}
\end{widetext}

Inserting Eq. \eqref{eq:4} and Eq. \eqref{eq:13} into Eq. \eqref{eq:12} gives for the optical chirality

\begin{align}
C &= \frac{\epsilon_0 \omega}{2c} \Big[\Bigl(\frac{k_z}{k} +\frac{k_t^2}{kk_z} +\frac{k_t^2k_z}{k^3}\Bigr)(J_{|\ell_\text{A}|}^2\sin^2\chi-J_{|\ell_\text{B}|}^2\cos^2\chi) \nonumber \\ \nonumber &+\frac{k_t^2}{kk_z}\Bigl( J_{|\ell_\text{A}|-1}^2\sin^2\chi -J_{|\ell_\text{B}|-1}^2\cos^2\chi \nonumber \\
&+ \Bigl(\frac{1}{2}+\frac{k_z^2}{2k^2}\Bigr)\Bigl(J_{|\ell_\text{A}|}J_{|\ell_\text{B}|-2}-J_{|\ell_\text{B}|}J_{|\ell_\text{A}|-2}\Bigr)\nonumber \\ &\times\cos((|\ell_\text{A}|+|\ell_\text{B}|-2)\phi+2\theta)\sin2\chi\Bigr)\Bigr]. 
\label{eq:14}
\end{align}

The spatial distributions of Eq.~\eqref{eq:14} are given in Fig. \ref{fig:3} for a varying range of $\ell_\text{P}$ and $m$. When $\chi = 0$ or $\pi/2$, Eq.~\eqref{eq:14} describes the optical chirality of a scalar right-circularly polarized with $\ell>0$ and left-circularly polarized with $\ell<0$ Bessel beam, respectively. Fig. \ref{fig:3} clearly exhibits that the optical chirality spatial distributions are identical for a given pair of $|\ell_\text{P}|$ and $|m|$, however the sign of the optical chirality density at a given location flips when changing the Pancharatnam charge handedness, i.e. $|\ell_\text{P}|$ versus $-|\ell_\text{P}|$. For a given $\ell_\text{P}$, the spatial distributions are highly-dependent on  $m$. Note, however, that the sign of $\ell_\text{P}$ does not influence the spatial distribution of the optical chirality. 

The final terms in Eq.~\eqref{eq:14} are interference terms between the zeroth-order and second-order transverse electromagnetic fields. These interference terms for scalar vortex beams are always zero for linearly polarized beams \cite{forbes2021measures, forbes2023customized}, though do have circularly symmetric non-zero contributions to the chirality for circularly-polarized modes. The additional, non-circularly symmetric (in general) terms of vector vortex beams we have highlighted here manifest due to the fact the second-order transverse electric (magnetic) fields of beam A have components which are $\pi/2$ out-of-phase with zeroth-order transverse magnetic (electric) fields of beam B (and vice versa). This phase relationship is unique to VVBs and clearly cannot manifest in a single scalar beam. Thus, the optical chirality Eq.~\eqref{eq:12}, which essentially measures the degree that the electric and magnetic field components are $\pi/2$ out of phase with one another yields a non-zero result for these interference terms. It is worth emphasizing a point not often made in the literature \cite{forbes2021measures} that when calculating optical properties of electromagnetic fields it is pivotal to include all contributions up to a given order of the paraxial parameter. Due to the fact the second-order transverse fields are an order smaller than the first-order longitudinal fields it is tempting to neglect them. However, when calculating a property which is the inner product of the fields (i.e. energy density and optical chirality density), the cross-terms between zeroth-order and second-order transverse fields are of the same general magnitude as the pure first-order longitudinal field contribution. For scalar beams the zeroth-order second-order interference terms generally add little qualitative difference to the spatial distributions, however this is clearly not true for vector beams as we have highlighted here.

\begin{figure*}
    \includegraphics[]{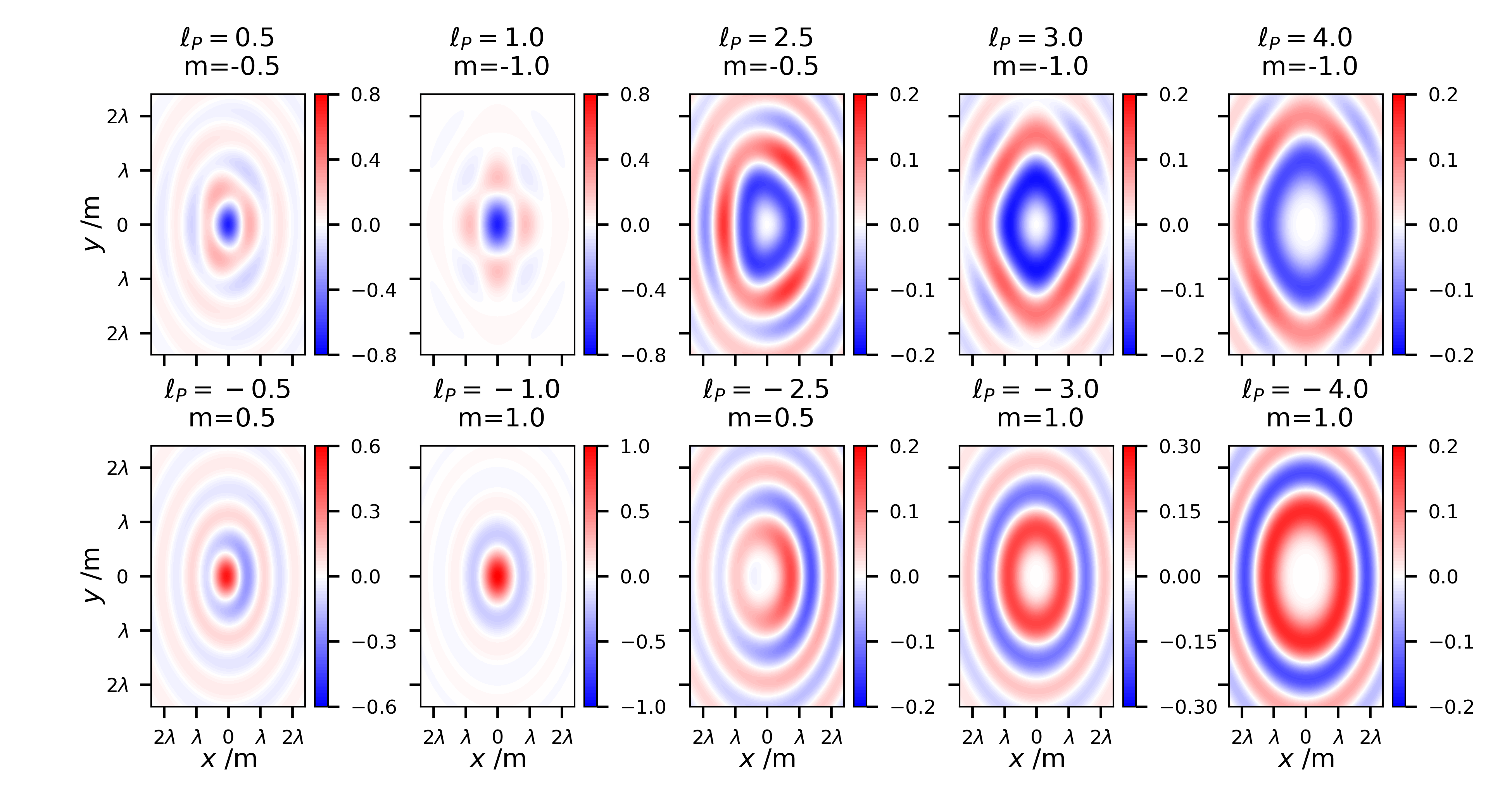}
    \caption{Optical chirality density spatial distributions of $ \ell_\text{A}^{\text{L}} + \ell_\text{B}^{\text{L}} $ vector beams Eq~.\eqref{eq:16}: Top row: $\ell_{\text{A}} > \ell_{\text{B}}$; Bottom: $\ell_{\text{A}} < \ell_{\text{B}}$, i.e. negative $\ell_{\text{P}}$ of Top row. }
    \label{fig:4}
\end{figure*}

\subsection{$\ell_\text{A}^{\text{L}} + \ell_\text{B}^{\text{L}}$ vector beams}

 We already have the required electric field which relates to the $\ell_\text{A}^{\text{L}} + \ell_\text{B}^{\text{L}}$ type CVVBs Eq. \eqref{eq:8}. In order to calculate the optical chirality density using Eq. \eqref{eq:12} we require the magnetic field. The corresponding magnetic field is extracted from Eq. \eqref{eq:2} as

\begin{widetext}
\begin{align}
\mathbf{B} &= \Bigl(J_{|\ell_\text{A}|}\text{e}^{i(|\ell_\text{A}|\phi-\theta)} \frac{k_z}{k} \sin\chi \begin{bmatrix}
-i\\
1
\end{bmatrix} 
+J_{|\ell_\text{B}|}\text{e}^{i(|\ell_\text{B}|\phi+\theta)} \frac{k_z}{k}\cos\chi \begin{bmatrix}
i\\ \nonumber
1
\end{bmatrix}  -\mathbf{\hat{z}}\frac{k_t}{k}\bigl(J_{|\ell_\text{A}|+1}\text{e}^{i[(|\ell_\text{A}|+1)\phi-\theta]}\sin\chi \nonumber \\ & + J_{|\ell_\text{B}|-1}\text{e}^{i[(|\ell_\text{B}|-1)\phi+\theta]}\cos\chi\bigl) + \frac{k_t^2}{2kk_z}\big[i\mathbf{\hat{x}}(-J_{|\ell_\text{A}|}\text{e}^{i(|\ell_\text{A}|\phi-\theta)}\sin\chi + J_{|\ell_\text{B}|}\text{e}^{i(|\ell_\text{B}|\phi+\theta)}\cos\chi \nonumber \\ \nonumber &- J_{|\ell_\text{A}|+2}\text{e}^{i[(|\ell_\text{A}|+2)\phi-\theta]}\sin\chi + J_{|\ell_\text{B}|-2}\text{e}^{i[(|\ell_\text{B}|-2)\phi+\theta]}\cos\chi) 
+\mathbf{\hat{y}}(J_{|\ell_\text{A}|}\text{e}^{i(|\ell_\text{A}|\phi-\theta)}\sin\chi + J_{|\ell_\text{B}|}\text{e}^{i(|\ell_\text{B}|\phi+\theta)}\cos\chi \nonumber \\  &- J_{|\ell_\text{A}|+2}\text{e}^{i[(|\ell_\text{A}|+2)\phi-\theta]}\sin\chi - J_{|\ell_\text{B}|-2}\text{e}^{i[(|\ell_\text{B}|-2)\phi+\theta]}\cos\chi)\bigr]\Bigr)\frac{1}{c}\text{e}^{ik_z z}.
\label{eq:15}
\end{align}
\end{widetext}

Inserting Eq. \eqref{eq:8} and Eq. \eqref{eq:15} into Eq. \eqref{eq:12} gives for the optical chirality

\begin{align}
C &= \frac{\epsilon_0 \omega}{2c} \Big[\Bigl(\frac{k_z}{k} +\frac{k_t^2}{kk_z} +\frac{k_t^2k_z}{k^3}\Bigr)(J_{|\ell_\text{A}|}^2\sin^2\chi-J_{|\ell_\text{B}|}^2\cos^2\chi) \nonumber \\ \nonumber &+\frac{k_t^2}{kk_z}\Bigl( J_{|\ell_\text{A}|+1}^2\sin^2\chi -J_{|\ell_\text{B}|-1}^2\cos^2\chi \nonumber \\
&+ \Bigl(\frac{1}{2}+\frac{k_z^2}{2k^2}\Bigr)\Bigl(J_{|\ell_\text{A}|}J_{|\ell_\text{B}|-2}-J_{|\ell_\text{B}|}J_{|\ell_\text{A}|+2}\Bigr)\nonumber \\ &\times\cos((|\ell_\text{A}|-|\ell_\text{B}|+2)\phi-2\theta)\sin2\chi\Bigr)\Bigr]. 
\label{eq:16}
\end{align}

The spatial distributions of Eq.~\eqref{eq:16} are given in Fig. \ref{fig:4} for a varying range of $\ell_\text{P}$ and $m$. When $\chi = 0$ or $\pi/2$, Eq~\eqref{eq:16} describes the chirality of a scalar right-circularly polarized with $\ell>0$ and left-circularly polarized with $\ell>0$ Bessel beam, respectively. In stark contrast to Fig. \ref{fig:3}, the sign of $\ell_\text{P}$ not only dictates the sign of optical chirality it also completely alters the spatial distributions. 

\section{Discussion and Conclusion}

Here we have provided an analytical description of the electromagnetic fields of cylindrical vector vortex beams using a co-propagating superposition of orthogonally polarized scalar Bessel beams. The equations can account for both a paraxial (weakly focused) or nonparaxial (tightly focused) CVVB. We highlighted the importance of including the second-order transverse electromagnetic field components for CVVBs. We used these electromagnetic fields to calculate the spin angular momentum density (both longitudinal and transverse) and optical chirality density for a variety of modes up to second-order in the paraxial parameter. We showed that CVVBs have rich and exotic spatial distributions of spin and chirality, and are highly tailorable through manipulating their Pancharatnam topological charge and polarization index. Using a general formulation of the 2D polarization state of the vector vortex beams, i.e. using the angles $\chi$ and $\theta$, including all possible combinations of $\ell_\text{A}$, $\ell_\text{B}$, $\sigma_\text{A}$, and $\sigma_\text{B}$ we produce analytical equations which can describe the optical spin angular momentum density and optical chirality density of each point on any higher-order or hybrid-order Poincar\'e sphere (in the circular polarization basis). The work can readily be extended to the elliptical or linear polarization bases for vector beams by using our general formulation, specifically Eqs.~\eqref{eq:1} and \eqref{eq:2}. Changing the basis will lead to different spatial distributions of the spin and chirality. Furthermore, it would be interesting to study real propagating modes that have a Gouy phase which should lead to interesting differences with the pure non-diffracting Bessel modes studied here, such as a rotation of the spatial patterns of spin and chirality along $z$. 

The rich spatial distributions of spin and chirality for VVBs we have highlighted can be utilized in spatially-dependent light-matter interactions \cite{wang2020vectorial, li2022transfer, aita2022enhancement}, exotic optical trapping landscapes for particles \cite{yang2021optical, otte2020optical, kritzinger2022optical}, alongside chiral sorting mechanisms \cite{forbes2022enantioselective, genet2022chiral, li2019optical, herrero2023combination}. Clear to see from the results of this work is that the handedness (or geometrical chirality) of the Pancharatnam topological charge $\ell_\text{P}$ (and thus phase) strongly influences the spatial distributions of the spin and chirality densities, even under weak focusing (paraxial) conditions. This is in stark contrast to the case of scalar vortex beams where the geometric chirality associated with $\ell$ does not influence the optical chirality density unless the scalar beam is under significant spatial confinement (tight focusing) \cite{forbes2021measures, forbes2023customized, green2023optical}. Finally it must be remembered that, due to spatial constraints, we have only provided simulations for a tiny fraction of possible variations in the given parameters which tailor CVVBs: amplitude, $\ell_{\text{P}}$, $m$, $\theta$, $\chi$, $k_t/k_z$, etc. Our simulations throughout this paper correspond to relatively low values of $\ell_{\text{P}}$ and $m$ at a single point on any given higher-order/hybrid-order Poincar\'e sphere. Higher values of $\ell_{\text{P}}$ and $m$ in varying combinations do yield even more complex spatial distributions: whilst these cannot be presented, our analytical equations do  account for this essentially infinite variation in the degrees of freedom.   

\section{Acknowledgements}

Dale Green is thanked for assistance with the production of the figures. David L. Andrews is thanked for comments. 

\section{Appendix I: $\ell_\text{A}^{\text{L}} + \ell_\text{B}^{\text{R}}$ beams}

The spin and chirality densities for $\ell_\text{A}^{\text{L}} + \ell_\text{B}^{\text{R}}$ type beams is given in this Appendix. For these beams $\ell_{\text{P}} = (|\ell_{\text{A}}|-|\ell_{\text{B}}|)/2$ and $m =-(|\ell_{\text{A}}|+|\ell_{\text{B}}||)/2$.

\begin{align}
{s^E_z} &= (J_{|\ell_\text{A}|}^2\sin^2\chi-J_{|\ell_\text{B}|}^2\cos^2\chi) (1+\frac{2k_t^2}{k^2}) \nonumber \\ \nonumber
&+\frac{k_t^2}{k^2}(J_{|\ell_\text{A}|}J_{|\ell_\text{B}|+2}-J_{|\ell_\text{B}|}J_{|\ell_\text{A}|+ 2}) \\ 
& \times \cos[(|\ell_\text{A}|+|\ell_\text{B}|+2)\phi-2\theta]\sin2\chi, 
\label{eq:17}
\end{align}

\begin{align}
{s^E_x} &= \frac{k_t}{k_z}\bigl[2(-J_{|\ell_\text{A}|}J_{|\ell_\text{A}|+1}\sin^2\chi -J_{|\ell_\text{B}|}J_{|\ell_\text{B}|+1}\cos^2\chi)\sin\phi \nonumber\\ 
&+(J_{|\ell_\text{B}|}J_{|\ell_\text{A}|+1} + J_{|\ell_\text{A}|}J_{|\ell_\text{B}|+1}) \nonumber \\ & \times \sin\bigl((|\ell_\text{A}|+|\ell_\text{B}|+1)\phi - 2\theta\bigr)\sin2\chi],
\label{eq:18}
\end{align}

and the $y$ component, 

\begin{align}
{s^E_y} &= \frac{k_t}{k_z}\bigl[2(J_{|\ell_\text{A}|}J_{|\ell_\text{A}|+1}\sin^2\chi +J_{|\ell_\text{B}|}J_{|\ell_\text{B}|+1}\cos^2\chi)\cos\phi \nonumber\\ 
&+(J_{|\ell_\text{B}|}J_{|\ell_\text{A}|+1} + J_{|\ell_\text{A}|}J_{|\ell_\text{B}|+1}) \nonumber \\ &\times \cos\bigl((|\ell_\text{A}|+|\ell_\text{B}|+1)\phi - 2\theta\bigr)\sin2\chi].
\label{eq:19}
\end{align}

\begin{figure*}
    \includegraphics[]{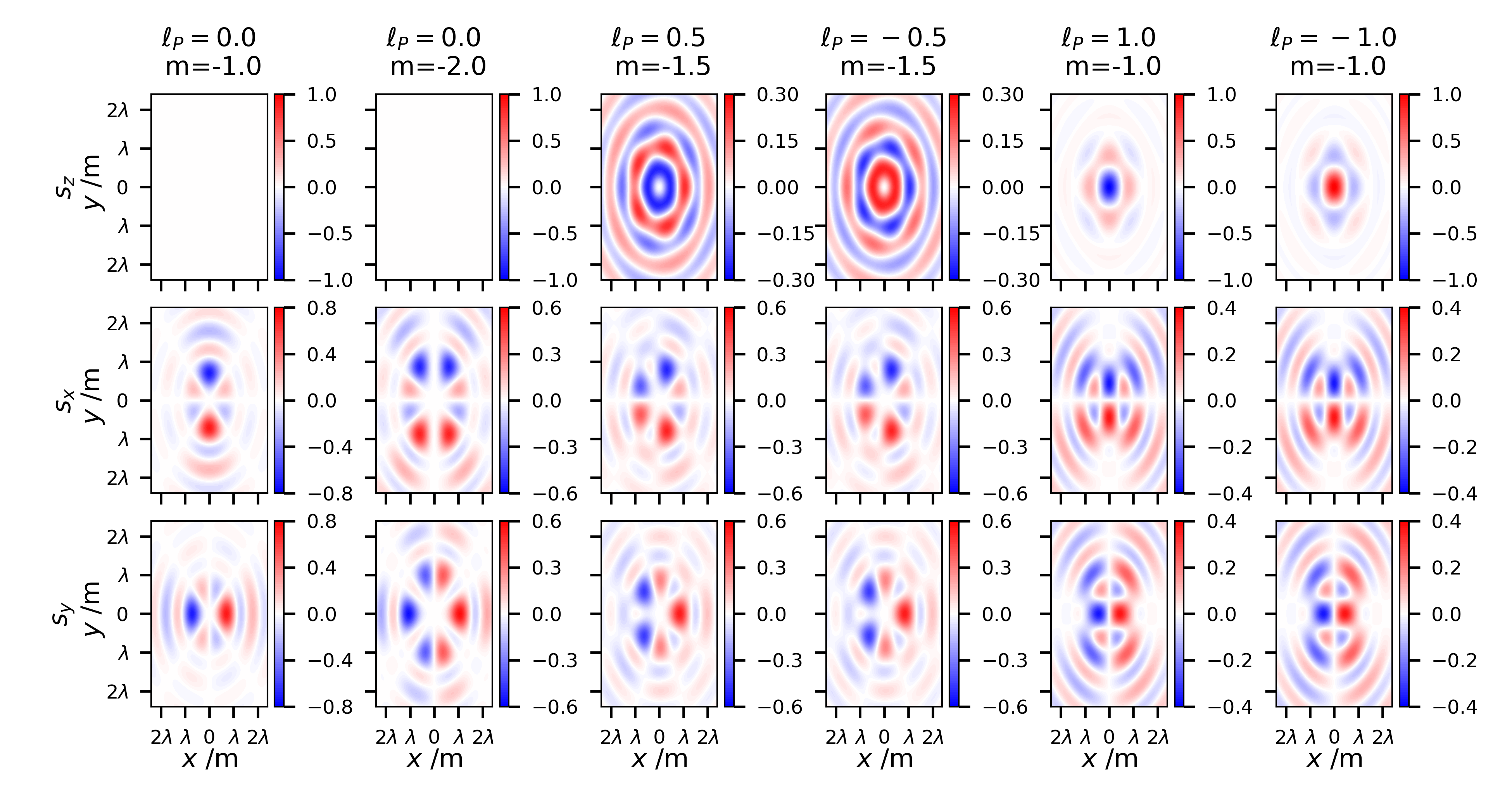}
    \caption{Spin angular momentum density spatial distributions of $ \ell_\text{A}^{\text{L}} + \ell_\text{B}^{\text{R}} $ vector beams: Top row: $z$-component Eq.~\eqref{eq:17}; Middle: $x$-component Eq.~\eqref{eq:18}; Bottom: $y$-component Eq.~\eqref{eq:19}.}
    \label{fig:5}
\end{figure*}

The spatial distributions of Eqs.~\eqref{eq:17}-\eqref{eq:19} are given in Fig. \ref{fig:5} for a varying range of $\ell_\text{P}$ and $m$. 

\begin{align}
C &= \frac{\epsilon_0 \omega}{2c} \Big[\Bigl(\frac{k_z}{k} +\frac{k_t^2}{kk_z} +\frac{k_t^2k_z}{k^3}\Bigr)(J_{|\ell_\text{A}|}^2\sin^2\chi-J_{|\ell_\text{B}|}^2\cos^2\chi) \nonumber \\ \nonumber &+\frac{k_t^2}{kk_z}\Bigl( J_{|\ell_\text{A}|+1}^2 \sin^2\chi -J_{|\ell_\text{B}|+1}^2 \cos^2\chi  \nonumber \\
&+ \Bigl(\frac{1}{2}+\frac{k_z^2}{2k^2}\Bigr)\Bigl(J_{|\ell_\text{A}|}J_{|\ell_\text{B}|+2}-J_{|\ell_\text{B}|}J_{|\ell_\text{A}|+ 2}) \nonumber \\ 
& \times \cos[(|\ell_\text{A}|+|\ell_\text{B}|+2)\phi-2\theta]\sin2\chi\Bigr)\Bigr] 
\label{eq:20}
\end{align}

\begin{figure*}
    \includegraphics[]{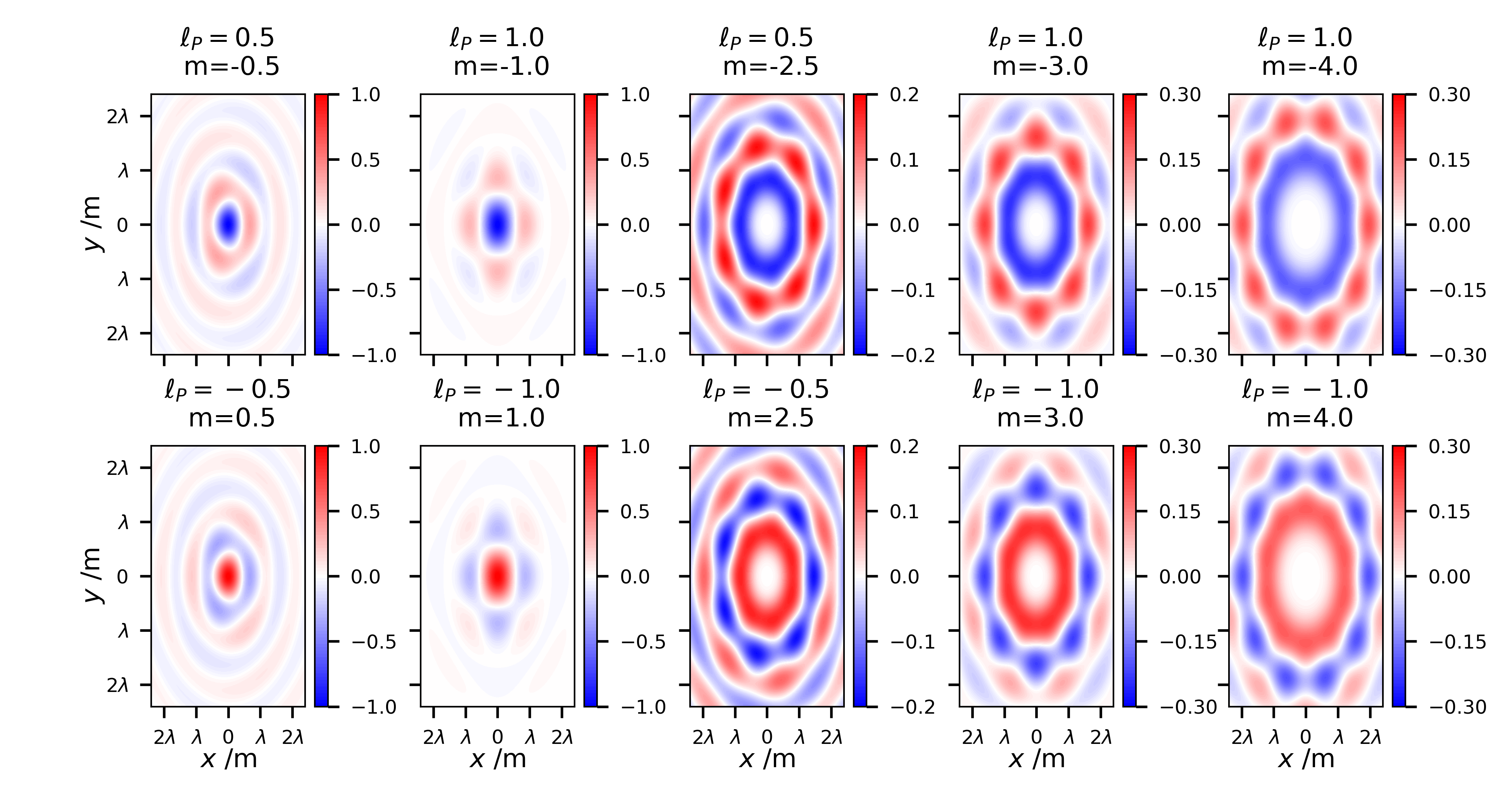}
    \caption{Optical chirality density spatial distributions of $ \ell_\text{A}^{\text{L}} + \ell_\text{B}^{\text{R}} $ vector beams Eq~.\eqref{eq:20}: Top row: $\ell_{\text{A}} > \ell_{\text{B}}$; Bottom: $\ell_{\text{A}} < \ell_{\text{B}}$, i.e. negative $\ell_{\text{P}}$ of Top row. }
    \label{fig:6}
\end{figure*}

The spatial distributions of Eq.~\eqref{eq:20} are given in Fig. \ref{fig:6} for a varying range of $\ell_\text{P}$ and $m$. 

\section{Appendix II: $\ell_\text{A}^{\text{R}} + \ell_\text{B}^{\text{R}}$ beams}

For these beams $ \text{sgn} |\ell_\text{A}| = \text{sgn} |\ell_\text{B}|$ where $\ell_\text{A/B} \leq 0 $  beams $\ell_{\text{P}} = -(|\ell_{\text{A}}|+|\ell_{\text{B}}|)/2$ and $m =(|\ell_{\text{A}}|-|\ell_{\text{B}}||)/2$.

The $z$ component of the spin angular momentum is

\begin{align}
{s^E_z} &= (J_{|\ell_\text{A}|}^2\sin^2\chi-J_{|\ell_\text{B}|}^2\cos^2\chi) (1+\frac{2k_t^2}{k^2}) \nonumber \\ \nonumber
&+\frac{k_t^2}{k^2}(J_{|\ell_\text{A}|}J_{|\ell_\text{B}|+2}-J_{|\ell_\text{B}|}J_{|\ell_\text{A}|-2}) \\ 
& \times \cos[(|\ell_\text{A}|-|\ell_\text{B}|-2)\phi+2\theta]\sin2\chi, 
\label{eq:21}
\end{align}

the $x$ component, 

\begin{align}
{s^E_x} &= \frac{k_t}{k_z}\bigl[2(J_{|\ell_\text{A}|}J_{|\ell_\text{A}|-1}\sin^2\chi -J_{|\ell_\text{B}|}J_{|\ell_\text{B}|+1}\cos^2\chi)\sin\phi \nonumber\\ 
&+(J_{|\ell_\text{B}|}J_{|\ell_\text{A}|-1}-J_{|\ell_\text{A}|}J_{|\ell_\text{B}|+1}) \nonumber \\ 
&\times \sin\bigl((|\ell_\text{A}|-|\ell_\text{B}|-1)\phi + 2\theta\bigr)\sin2\chi],
\label{eq:22}
\end{align}

and the $y$ component, 

\begin{align}
{s^E_y} &= -\frac{k_t}{k_z}\bigl[2(J_{|\ell_\text{A}|}J_{|\ell_\text{A}|-1}\sin^2\chi -J_{|\ell_\text{B}|}J_{|\ell_\text{B}|+1}\cos^2\chi)\cos\phi \nonumber\\ 
&+(J_{|\ell_\text{B}|}J_{|\ell_\text{A}|-1}-J_{|\ell_\text{A}|}J_{|\ell_\text{B}|+1}) \nonumber \\ 
&\times \cos\bigl((|\ell_\text{A}|-|\ell_\text{B}|-1)\phi + 2\theta\bigr)\sin2\chi].
\label{eq:23}
\end{align}

\begin{figure*}
    \includegraphics[]{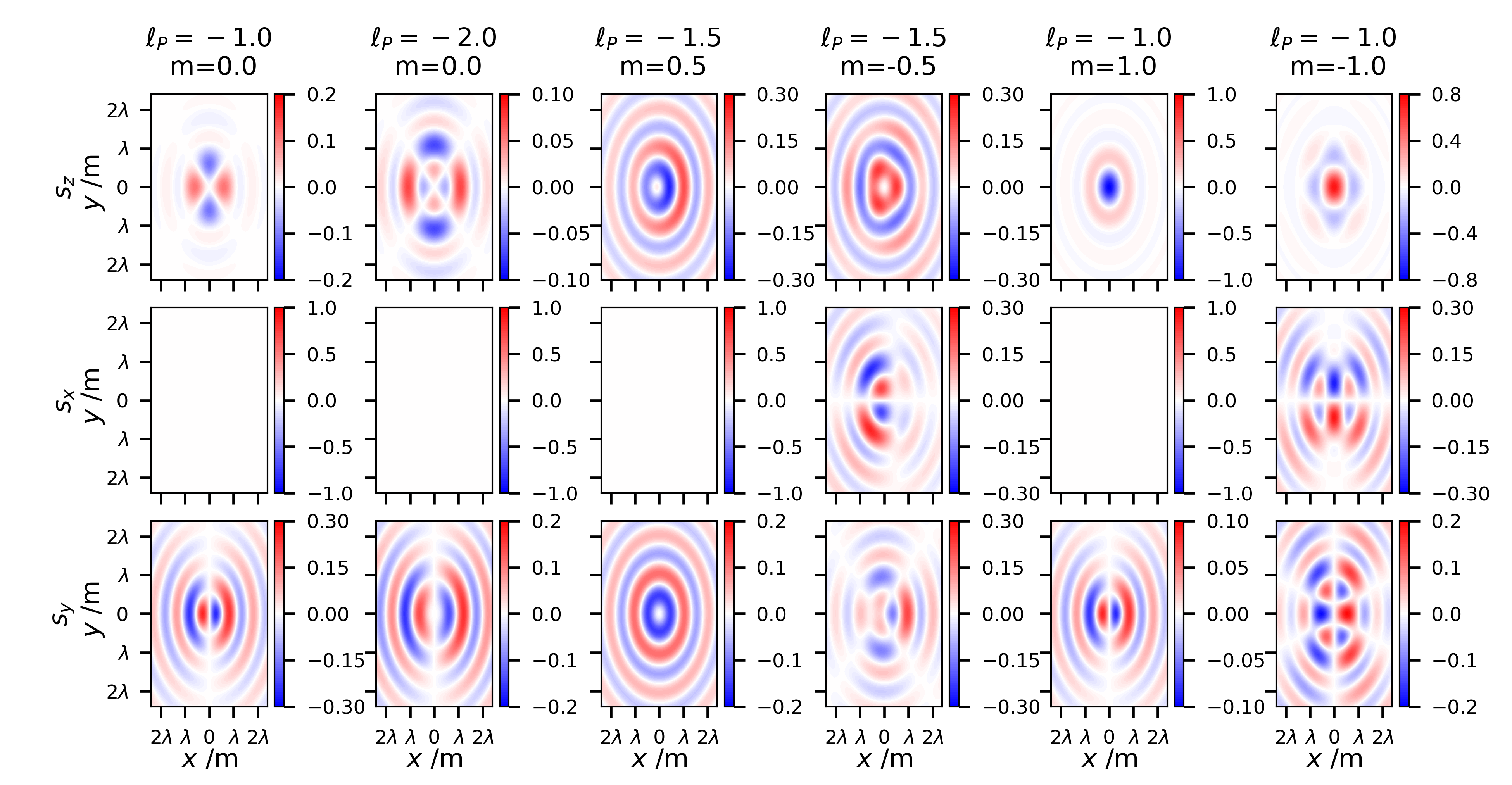}
    \caption{Spin angular momentum density spatial distributions of $ \ell_\text{A}^{\text{R}} + \ell_\text{B}^{\text{R}} $ vector beams: Top row: $z$-component Eq.~\eqref{eq:21}; Middle: $x$-component Eq.~\eqref{eq:22}; Bottom: $y$-component Eq.~\eqref{eq:23}.}
    \label{fig:7}
\end{figure*}

The spatial distributions of Eqs.~\eqref{eq:21}-\eqref{eq:23} are given in Fig. \ref{fig:7} for a varying range of $\ell_\text{P}$ and $m$. 

The optical chirality density is

\begin{align}
C &= \frac{\epsilon_0 \omega}{2c} \Big[\Bigl(\frac{k_z}{k} +\frac{k_t^2}{kk_z} +\frac{k_t^2k_z}{k^3}\Bigr)(J_{|\ell_\text{A}|}^2\sin^2\chi-J_{|\ell_\text{B}|}^2\cos^2\chi) \nonumber \\ \nonumber &+\frac{k_t^2}{kk_z}\Bigl( J_{|\ell_\text{A}|-1}^2\sin^2\chi -J_{|\ell_\text{B}|+1}^2\cos^2\chi \nonumber \\
&+ \Bigl(\frac{1}{2}+\frac{k_z^2}{2k^2}\Bigr)\Bigl(J_{|\ell_\text{A}|}J_{|\ell_\text{B}|+2}-J_{|\ell_\text{B}|}J_{|\ell_\text{A}|-2}\Bigr)\nonumber \\ &\times\cos((|\ell_\text{A}|-|\ell_\text{B}|-2)\phi+2\theta)\sin2\chi\Bigr)\Bigr] 
\label{eq:24}
\end{align}

\begin{figure*}
    \includegraphics[]{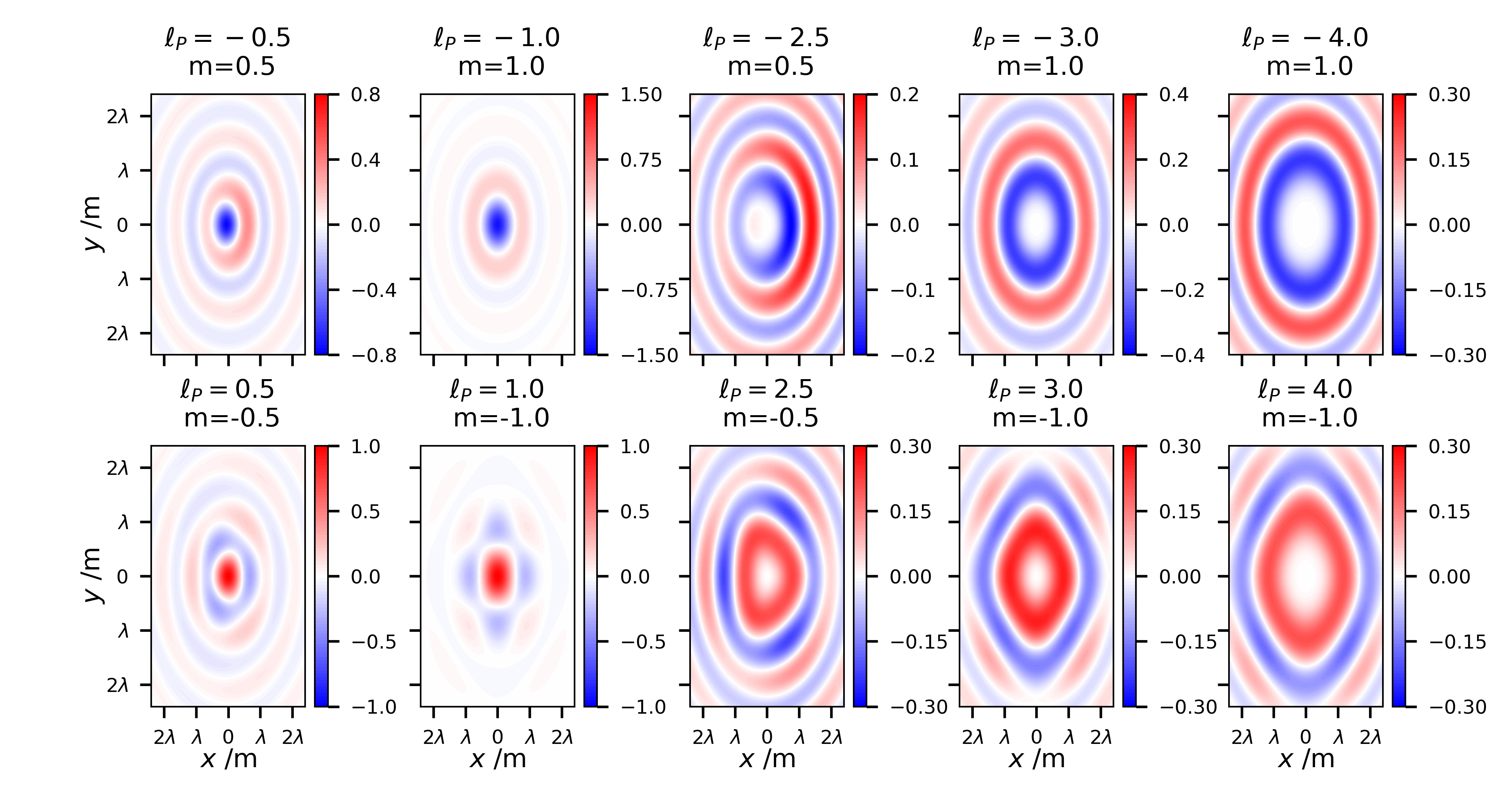}
    \caption{Optical chirality density spatial distributions of $ \ell_\text{A}^{\text{R}} + \ell_\text{B}^{\text{R}} $ vector beams Eq~.\eqref{eq:24}: Top row: $\ell_{\text{A}} > \ell_{\text{B}}$; Bottom: $\ell_{\text{A}} < \ell_{\text{B}}$, i.e. negative $\ell_{\text{P}}$ of Top row. }
    \label{fig:8}
\end{figure*}

The spatial distributions of Eq.~\eqref{eq:24} are given in Fig. \ref{fig:8} for a varying range of $\ell_\text{P}$ and $m$. 

\section{Appendix III: electromagnetic fields of a scalar Bessel beam}

Here we discuss the derivation of the electromagnetic fields for scalar Bessel beams. The form of the electromagnetic fields of a Bessel beam found throughout the literature (e.g. \cite{vicente2021bessel, quinteiro2019reexamination, forbes2023customized, bliokh2010angular, volke2002orbital, andrews2012angular}) can describe the optical properties (energy, momentum, optical chirality) of scalar beams by virtue of the fact all of these quantities are quadratic in the fields. However, when attempting to use scalar Bessel beams to construct vector beams an issue became apparent, namely the forms currently found in literature cannot achieve this (note \cite{ornigotti2013radially} use a description which does achieve it but not via the superposition of two scalar beams). There are two important points which authors have previously neglected (including the author of this paper): 1) The Bessel function order must be the absolute value of the topological charge $J_{|\ell|}$, not the topological charge $J_\ell$ 2) when deriving higher-order fields, the Bessel relations have to be used in a consistent way with aspect 1). Firstly it is worth giving an explicit example of the most obvious issue of trying to construct a simple first-order radially polarized beam using two scalar Bessel modes in the form found throughout the literature. The zeroth-order electric field for a  Bessel beam is (almost always) given in the literature as:

\begin{align}
\mathbf{E}^\text{T0}= (\alpha \mathbf{\hat{x}} + \beta \mathbf{\hat{y}}) J_{{\ell}}[k_tr]\text{e}^{i({k_z}z+\ell\phi)}
\label{eq:25}
\end{align}

To construct a first-order radially-polarized beam we require the superposition of a left-circularly polarized mode with $\ell = -1$ and a right-circularly polarized mode with $\ell = 1$: 

\begin{align}
&\frac{1}{\sqrt{2}}(\mathbf{\hat{x}} + i \mathbf{\hat{y}}) J_{-1}\text{e}^{i({k_z}z-\phi)} + \frac{1}{\sqrt{2}}(\mathbf{\hat{x}} - i \mathbf{\hat{y}}) J_{1}\text{e}^{i({k_z}z+\phi)} \nonumber \\
&= \frac{1}{\sqrt{2}}\mathbf{\hat{r}}(J_{-1}+J_1)\text{e}^{i{k_z z}} \nonumber \\ 
&= 0
\label{eq:26}
\end{align}

Because $J_{-1}=-J_1$, The correct zeroth-order transverse field of Bessel beam is given as

\begin{align}
\mathbf{E}^\text{T0}= (\alpha \mathbf{\hat{x}} + \beta \mathbf{\hat{y}}) J_{|{\ell}|}[k_tr]\text{e}^{i({k_z}z+\ell\phi)}
\label{eq:27}
\end{align}

Note the fact the Bessel mode has the order of the modulus of topological charge. With this small change implemented and reflected in Eq. \eqref{eq:18} then the first-order radially polarized field is readily constructed. This small correction has much deeper consequences when the higher-order longitudinal and transverse fields are required for a tight focussed Bessel mode. With the aid of Gauss's Law, it is simple to show the first-order longitudinal field component of a scalar Bessel beam is 

\begin{align}
E^\text{L1}_z= \frac{i}{k_z}\Bigl(\alpha \frac{\partial }{\partial x} + \beta \frac{\partial }{\partial y} \Bigr) J_{|{\ell}|}[k_tr]\text{e}^{i({k_z}z+\ell\phi)}
\label{eq:28}
\end{align}

Converting the Cartesian partial derivatives to cylindrical coordinates we readily come to the expression:

\begin{align}
E^\text{L1}_z &= \frac{i}{k_z}\Bigl(\alpha\Bigl[ \cos\phi\Bigl\{\frac{k_t}{2}(J_{|\ell|-1} - J_{|\ell|+1})\Bigr\} - \frac{i\ell}{r}J_{|\ell|}\sin\phi\Bigr] \nonumber \\ \nonumber
& + \Bigl(\beta\Bigl[ \sin\phi\Bigl\{\frac{k_t}{2}(J_{|\ell|-1} - J_{|\ell|+1})\Bigr\} + \frac{i\ell}{r}J_{|\ell|}\cos\phi\Bigr] \\ 
& \times \text{e}^{i({k_z}z+\ell\phi)}
\label{eq:29}
\end{align}

In the derivations used so far in the literature the two terms linearly dependent on $\ell$ are given as (remembering most authors neglect the modulus of the order of the Bessel function): 

\begin{align}
-\frac{i\ell}{r}J_{\ell}\sin\phi  
\label{eq:30}
\end{align}
\begin{align}
 \frac{i\ell}{r}J_{\ell}\cos\phi 
\label{eq:31}
\end{align}

At this point the following Bessel relation is used:

\begin{align}
\frac{2\ell}{k_tr}J_\ell[k_tr] = J_{\ell+1}[k_tr]+J_{\ell-1}[k_tr]
\label{eq:32}
\end{align}

which leads to following form of the longitudinal component 

\begin{align}
E^\text{L1}_z &= \frac{ik_t}{2k_z}\Bigl((\alpha+i\beta)J_{\ell-1}\text{e}^{-i\phi} + (i\beta-\alpha)J_{\ell+1}\text{e}^{i\phi}\Bigr) \nonumber\\ 
& \times \text{e}^{i({k_z}z+\ell\phi)}
\label{eq:33}
\end{align}

However, what we actually have is 

\begin{align}
-\frac{i\ell}{r}J_{|\ell|}\sin\phi = \mp\frac{i|\ell|}{r}J_{|\ell|}\sin\phi 
\label{eq:34}
\end{align}
\begin{align}
 \frac{i\ell}{r}J_{|\ell|}\cos\phi =
\pm\frac{i|\ell|}{r}J_{|\ell|}\cos\phi 
\label{eq:35}
\end{align}

In which the correct Bessel relation to use is:

\begin{align}
\frac{2|\ell|}{k_tr}J_{|\ell|}[k_tr] = J_{|\ell|+1}[k_tr]+J_{|\ell|-1}[k_tr]
\label{eq:36}
\end{align}

which leads to following correct form of the longitudinal component 

\begin{align}
E^\text{L1}_z &= \frac{ik_t}{2k_z}\Bigl((\alpha\pm i\beta)J_{|\ell|-1}\text{e}^{\mp i\phi} + (i\beta\mp\alpha)J_{|\ell|+1}\text{e}^{\pm i\phi}\Bigr) \nonumber\\ 
& \times \text{e}^{i({k_z}z+\ell\phi)}
\label{eq:37}
\end{align}

where the upper sign corresponds to $\ell>0$ and lower sign $\ell<0$: for $\ell=0$ the sign does not matter, both give equivalent and correct results. Following similar algebra the second-order transverse fields can be derived. The correct formula for Bessel beams which should be used from now on are Eqs. \eqref{eq:1} and \eqref{eq:2} of the main manuscript.

\bibliographystyle{apsrev4-1}
\bibliography{references.bib}
\end{document}